# Inverse and direct energy cascades in 3D MHD turbulence at low $Rm$

Nathaniel T. Baker,[1,2,3] Alban Pothérat,[1] Laurent Davoust,[3] and François Debray[2]
[1] *Coventry University, Applied Mathematics Research Centre, Coventry CV15FB, UK*
[2] *LNCMI-EMFL-CNRS, UGA, INSA, UPS 25 Ave. des Martyrs, 38000 Grenoble, France*
[3] *Grenoble-INP/CNRS/Univ. Grenoble-Alpes, SIMaP EPM, F-38000 Grenoble, France*

(Dated: 8 June 2018)

This experimental study analyzes the relationship between the dimensionality of turbulence and the up- or downscale nature of its energy transfers. We do so by forcing low-$Rm$ magnetohydrodynamic (MHD) turbulence in a confined channel, while precisely controlling its dimensionality by means of an externally applied magnetic field. We first identify a specific lengthscale $\hat{l}_\perp^c$ that separates smaller 3D structures from larger quasi-2D ones. We then show that an inverse energy cascade of horizontal kinetic energy along horizontal scales is always observable at large scales, but that it extends well into the region of 3D structures. At the same time, a direct energy cascade confined to the smallest and strongly 3D scales is observed. These dynamics therefore appear not to be simply determined by the dimensionality of individual scales, nor by the forcing scale, unlike in other studies. In fact, our findings suggest that the relationship between kinematics and dynamics is not universal and may strongly depend on the forcing and dissipating mechanisms at play.

Turbulence displays radically opposite dynamics, whether it is three-dimensional (3D) or two-dimensional (2D). In the former, kinetic energy follows a direct energy cascade from the forcing scale down to the smallest scales[1], while the latter features an inverse energy cascade from the forcing scale up to large structures of the size of the system[2]. It is, however, still unclear how these seemingly irreconcilable dynamics relate to each other, whenever 2D and 3D turbulent structures coexist. This question is all the more crucial when dealing with real-life wall-bounded flows, as speaking of two-dimensionality only makes sense with respect to the presence of boundaries, such as no-slip walls. Yet, solid boundaries necessarily introduce three-dimensionality both in boundary layers and in the bulk[3,4]. As a result, real flows (such as oceans or atmospheres) can only be quasi-2D rather than strictly 2D, and often combine 2D and 3D turbulent structures[5]. The key question that determines both transport and dissipative properties of such flows is then how much, and which kind of three-dimensionality is required for the inverse cascade to become direct. In other words: how do the energy transfers relate to the topological dimensionality of turbulence?

It is unclear whether this question has a universal answer. For instance, compressing one dimension can yield a hybrid configuration, in which the energy flux splits into a direct cascade at small scales and an inverse cascade at large scales[6], while forcing a 3D and three-component flow in a thick fluid layer can still produce a large coherent vortex, indicative of an upscale energy flux[7]. Furthermore, within the respective contexts of rotating[8], and stratified rotating quasi-2D turbulence[9], horizontal kinetic energy flows preferentially upscale, while vertical kinetic energy flows downscale. Finally, a subset of the non-linear interactions of any 3D flow is always capable of transferring kinetic energy upscale[10].

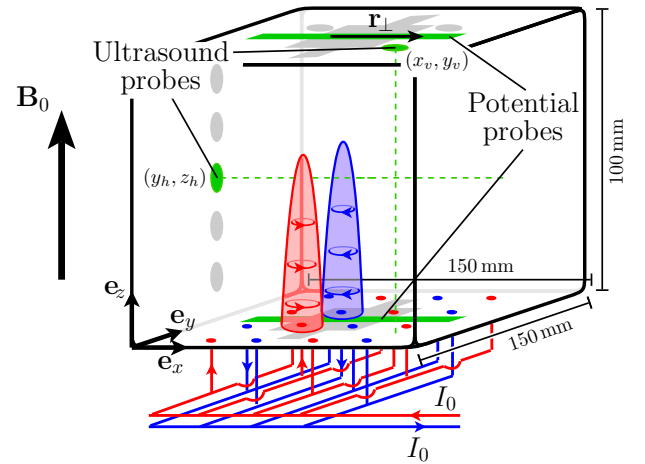

FIG. 1: Sketch of the Flowcube.

This matter is investigated within the context of statistically steady liquid metal low-$Rm$ magnetohydrodynamic (MHD) turbulence in a homogeneous magnetic field[12–14]. A significant advantage of this approach is that the level of three-dimensionality of MHD turbulence can be controlled simply by adjusting the external magnetic field $\mathbf{B}_0$[15–18]. In particular, Ref.[19] theorised that a critical lengthscale separates (larger) quasi-2D from (smaller) 3D turbulent structures, by interpreting the effect of the solenoidal component of the Lorentz force as a "pseudo-diffusion" of momentum in the direction of the magnetic field. The time $\tau_{2D}(l_\perp)$ required to diffuse the momentum of a turbulent structure of size $l_\perp$ over the distance $l_z$ along $\mathbf{B}_0$ is given by $\tau_{2D} = (\rho/\sigma B_0^2)(l_z/l_\perp)^2$, where $\sigma$ and $\rho$ are the fluid's electric conductivity and density respectively. In the inertial range, the other competing process is inertia, whose main effect is to redistribute kinetic energy across turbulent structures, by means of energy transfers. It takes place over the eddy turnover time $\tau_u(l_\perp) = l_\perp/u(l_\perp)$, where $u(l_\perp)$ is the velocity of the structure at hand. The scaling law for the range of action of the Lorentz force follows from the balance between both effects[19]:

$$l_z(l_\perp) = l_\perp \sqrt{N(l_\perp)}, \qquad (1)$$

where $N(l_\perp, u(l_\perp)) = \sigma B_0^2 l_\perp/\rho u(l_\perp)$ is a scale-dependent interaction parameter. The dimensionality of a structure is then determined with respect to no-slip walls perpendicular to the magnetic field and distant by $h$, through the ratio $l_z(l_\perp)/h$[20,21]. In particular, $l_z(l_\perp)/h \leqslant 1$ implies that velocity gradients exist in the bulk, in other words, that the structure of size $l_\perp$ is 3D. Conversely, $l_z(l_\perp)/h \gg 1$ implies that the Lorentz force diffuses the momentum of the structure of size $l_\perp$ over a distance much greater than $h$. This process is however blocked by the no slip-walls. The structure of size $l_\perp$ is thus quasi-2D. The critical lengthscale $l_\perp^c$ separating quasi-2D and 3D structures, for which $l_z(l_\perp^c)/h = 1$, eventually reads[19]

$$\frac{l_\perp^c}{h} \sim \left[\frac{\sigma B_0^2 h}{\rho u(l_\perp^c)}\right]^{-1/3} = N[h, u(l_\perp^c)]^{-1/3}. \qquad (2)$$

| | $l_i = 5\,\text{mm}$ | | | | | | $l_i = 15\,\text{mm}$ | | | | |
|---|---|---|---|---|---|---|---|---|---|---|---|
| $B_0$ [T] | 1 | 3 | 5 | 7 | 10 | $B_0$ [T] | 1 | 3 | 5 | 7 | 10 |
| $u_{\text{bot}}$ [m/s] | 0.180 | 0.230 | 0.240 | 0.250 | 0.270 | $u_{\text{bot}}$ [m/s] | 0.130 | 0.180 | 0.200 | 0.230 | 0.250 |
| $Ha$ | 3644 | 10930 | 18220 | 25510 | 36440 | $Ha$ | 3644 | 10930 | 18220 | 25510 | 36440 |
| $Re$ | 44000 | 58000 | 60000 | 64000 | 67000 | $Re$ | 32000 | 45000 | 50000 | 57000 | 62000 |
| $l_z(l_i)/h$ | 0.23 | 0.59 | 0.97 | 1.3 | 1.7 | $l_z(l_i)/h$ | 1.3 | 3.4 | 5.3 | 6.9 | 9.4 |

TABLE I: Range of non-dimensional parameters for an injected current per electrode of 6 A. Data is given for both electrode separations $l_i$, and a selected range of magnetic fields (cf. Ref.[11] for more details).

Increasing the applied magnetic field thus offers a convenient way of broadening the spectrum of quasi-2D scales.

The problem at hand was tackled experimentally using the Flowcube[11,20,22,23], an experimental platform designed to drive turbulence electrically in a $100\,\text{mm} \times 100\,\text{mm} \times 150\,\text{mm}$ parallelipedic vessel, filled with a GaInSn eutectic alloy ($\rho = 6400\,\text{kg/m}^3$, $\sigma = 3.4 \times 10^6\,\text{S/m}$, kinematic viscosity $\nu = 4 \times 10^{-7}\,\text{m}^2/\text{s}$). Turbulent motions were induced by forcing a DC electric current $I_0$ through a square periodic array of electrodes spaced either by $l_i = 5$ or $15\,\text{mm}$ located along the bottom wall[11], while simultaneously applying a vertical and static magnetic field $B_0\,\mathbf{e}_z$, of up to 10 T (cf. Fig. 1). Two complementary measurement methods were used to diagnose the resulting flow. First, a fine Cartesian mesh of probes mounted flush to the top and bottom walls along strips aligned with the $\mathbf{e}_x$ direction gave access to the electric potential distribution at these walls. The spatial resolution of this method, as given by the spacing between adjacent probes was 2.5 mm. The signal was time sampled at 250 Hz/24-bits over 18 mn-long continuous runs. In the limit of high Hartmann number ($Ha = B_0 h \sqrt{\sigma/\rho\nu}$) and high interaction parameter, the electric potential along the horizontal walls is a precise estimate for the stream-function right outside the Hartmann boundary layer developing along them[24]. It thus provides both velocity components in the same planes. Second, two ultrasound transducers were used to respectively measure the $u_x(x, y_h, z_h, t)$ and $u_z(x_v, y_v, z, t)$ velocity profiles through the bulk, at the fixed positions $(y_h, z_h)$ and $(x_v, y_v)$ respectively. The transducers offered a spatial resolution of ca. 1 mm, and a 5 Hz sampling rate. The dimensionality of the flow was controlled through the single parameter $l_z(l_i)/h$, where $l_z(l_i)$ is the diffusion length associated to turbulent structures of size $l_i$ and the RMS of the turbulent fluctuations measured along the bottom wall $u_{\text{bot}}$[11]. The Reynolds number $Re = u_{\text{bot}} h/\nu$ ranged between 17000 and 71000 throughout, which guaranteed that the turbulence was fully developed. The Hartmann number and Magnetic Reynolds number $Rm = \mu_0\,\sigma\,u_{\text{bot}}\,h$, ($\mu_0$ referring to vacuum permitivity) respectively ranged between 911 and 36400 and 0.029 and 0.12. Selected regimes achievable by the Flowcube are given in Table I. Except for Fig 4, the statistics presented hereafter stem from data acquired by potential measurements. The average operator must be understood as an ensemble average obtained by averaging over time and space. Statistics were computed using ca. $10^7$ independent samples, which yielded a convergence level better than 1% for the third order moments[11].

Following Ref.[25], we describe the structure of turbulence through the velocity increment $\delta\mathbf{u} = \mathbf{u}(\mathbf{x}+\mathbf{r}) - \mathbf{u}(\mathbf{x})$, computed from turbulent fluctuations. Due to the very low influence from the lateral walls[11], the turbulence in Flowcube is considered homogeneous in the horizontal plane and axisymmetric. Hence, $\mathbf{r}$ and $\delta\mathbf{u}$ are respectively decomposed as $\mathbf{r} = r_\perp\,\mathbf{e}_r + r_\parallel\,\mathbf{e}_z$, and $\delta\mathbf{u} = \delta\mathbf{u}_\perp + \delta\mathbf{u}_\parallel$, with $\delta\mathbf{u}_\parallel = (\delta\mathbf{u} \cdot \mathbf{e}_z)\,\mathbf{e}_z$ and $\delta\mathbf{u}_\perp = \delta\mathbf{u} - \delta\mathbf{u}_\parallel$. We shall focus on $\delta\mathbf{u}_\perp(r_\perp\,\mathbf{e}_x)$ computed along both the top and bottom plates.

Let us start by analyzing the kinematics of the turbulence and attempt to discriminate 3D from quasi-2D structures. To do so, we adopt the signature function

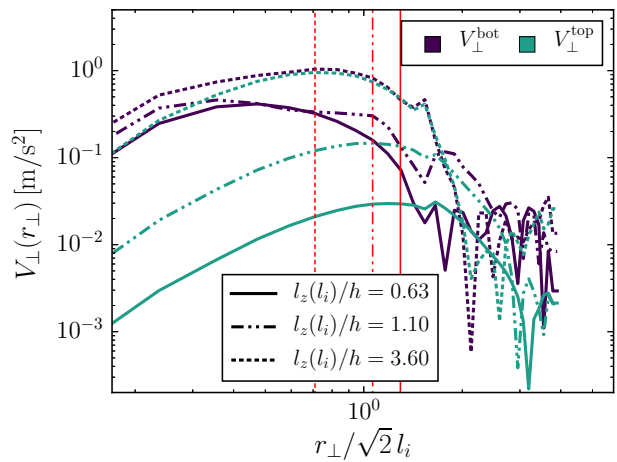

FIG. 2: Scale-wise perpendicular energy density along horizontal scales. Normalization by $\sqrt{2}$ is introduced so that the abscissae of Figs. 2 and 5 coincide with each other (cf. properties of $V_\perp$ in Ref[27]). The vertical red lines locate $\hat{l}_\perp^c/l_i$ for each $l_z(l_i)/h$, with quasi-2D scales lying to their right.

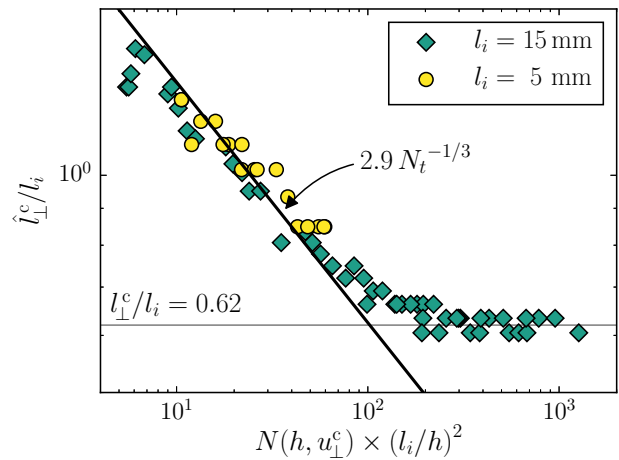

FIG. 3: 3D to quasi-2D critical lengthscale $\hat{l}_\perp^c$, as a function of the "true" interaction parameter $N_t = N(h, u_\perp^c) \times (l_i/h)^2$. The $N_t^{-1/3}$ region proves that the threshold between 3D and quasi-2D structures is indeed solely controlled by a balance between the solenoidal component of the Lorentz force and inertia.

$V$ as a scale-space alternative to the Fourier-space 3D energy spectrum[26], which is expressed in 2D as[27]

$$V_\perp(r_\perp) = -\frac{r_\perp^2}{4}\frac{\partial}{\partial r_\perp}\frac{1}{r_\perp}\frac{\partial\langle\delta u_l^2\rangle}{\partial r_\perp}. \quad (3)$$

Here, $\delta u_l = [\mathbf{u}(\mathbf{x} + r_x\mathbf{e}_x) - \mathbf{u}(\mathbf{x})] \cdot \mathbf{e}_x$ is the longitudinal velocity increment measured in the horizontal plane. In axisymmetric turbulence, quasi-2D structures are invariant with respect to $z$ outside the boundary layers. Their signature function must therefore be the same whether measured along the top or bottom walls. Conversely, any departure from a top/bottom mirror symmetry is an indication of a 3D structure. Fig. 2 shows the scale-wise distribution of $V_\perp(r_\perp)$ across horizontal structures, along the top and bottom walls (referred to as $V_\perp^{\text{top}}$ and $V_\perp^{\text{bot}}$ respectively). As $l_z(l_i)/h$ increases beyond one, $V_\perp^{\text{top}}$ tends to match $V_\perp^{\text{bot}}$, both in shape and amplitude. Based on this observation, a lower bound for the smallest quasi-2D scale is computed from the location of $V_\perp^{\text{top}}(r_\perp)$'s maximum $\hat{l}_\perp^c$. Interestingly, the superposition of top and bottom energy distributions starts at large scales and works its way through smaller and smaller scales as $l_z(l_i)/h$ increases. This behavior is in full agreement with Eq. (1), which states that it takes a higher field [i.e. a higher $l_z(l_i)/h$] to make smaller structures quasi-2D. Furthermore, the critical lengthscale $\hat{l}_\perp^c$ strikingly coincides with the scale at which $V_\perp^{\text{bot}}$ and $V_\perp^{\text{top}}$ depart from each other, thus confirming its physical relevance.

Fig. 3 reports the variations of $\hat{l}_\perp^c/l_i$ for all operating conditions explored against the "true" interaction parameter $N_t = N(h, u_\perp^c) \times (l_i/h)^2 = [l_z(l_i)/h]^2 (h/l_i)$, which measures the ratio of diffusion by the Lorentz force to inertia at the forcing scale[28]. Here, $u_\perp^c = [2 V_\perp^{\text{top}}(\hat{l}_\perp^c)\hat{l}_\perp^c]^{1/2}$ is an estimate for the velocity at scale $\hat{l}_\perp^c$. All measurements collapse onto a single curve, of which two parts can be singled out. For $N_t \lesssim 10^2$, $\hat{l}_\perp^c/l_i \propto N_t^{-1/3}$, which provides an experimental confirmation of Eq. (2). For $N_t \gtrsim 10^2$, $\hat{l}_c/l_i$ saturates towards a constant value of 0.62, indicating that scales below this limit-size cannot be quasi-2D no matter how high $N_t$ might be. This limit likely results from the absence of a mechanism to transfer energy to 2D scales smaller than the energy injection scale, a phenomenon which is not accounted for in (2).

Having identified quasi-2D and 3D regions of the scale space, we now seek regions where energy is transferred up- and downscale. We first recall that the equation governing energy transfers in statistically steady MHD turbulence is the Kàrmàn-Howarth equation, which reads, in the inhomogeneous and anisotropic case[29]

$$\Pi(\mathbf{r}) = \mathcal{P}(\mathbf{r}) + \mathcal{T}(\mathbf{r}) - \epsilon_J(\mathbf{r}) - \epsilon_\nu(\mathbf{r}). \quad (4)$$

$\Pi(\mathbf{r}) = \nabla_\mathbf{r} \cdot \langle|\delta\mathbf{u}|^2 \delta\mathbf{u}\rangle$ quantifies the flux of turbulent kinetic energy in scale space, $\mathcal{P}(\mathbf{r})$ is the rate of production of turbulent kinetic energy, $\mathcal{T}(\mathbf{r})$ is the flux of turbulent kinetic energy in physical space (due to spatial inhomogeneities), $\epsilon_J$ is the Joule dissipation (occurring in the bulk and the Hartmann layers[13]), while $\epsilon_\nu$ represents viscous dissipation. In low-$Rm$ MHD, the energy transfers remain confined to the usual non-linear hydrodynamic term $\Pi(\mathbf{r})$, which represents a local cumulative flux of kinetic energy exchanged between scales of size $r = \|\mathbf{r}\|$ and less, with those of size $r$ and greater[8]. More specifically, $\Pi(r) > 0$ (resp. $\Pi(r) < 0$) implies that, on average, energy flows towards scales larger (resp. smaller) than $r$, i.e. following an inverse (resp. direct) energy cascade.

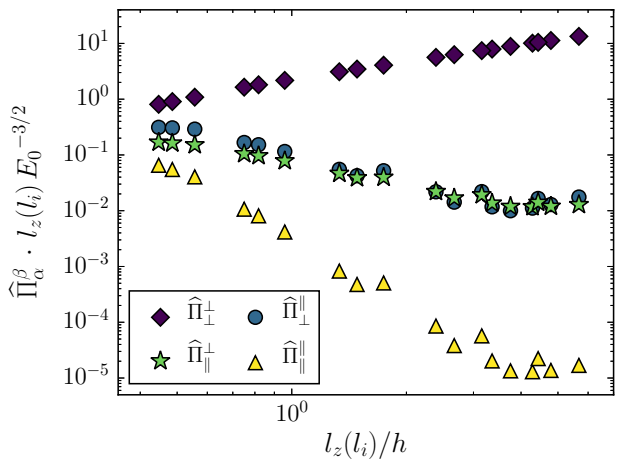

FIG. 4: Global estimates for the different contributions to $\Pi$, computed from ultrasound measurements. Regardless of the dimensionality of the flow, the main contribution to the cascade is $\Pi_\perp^\perp$, which can accurately be measured using potential probes.

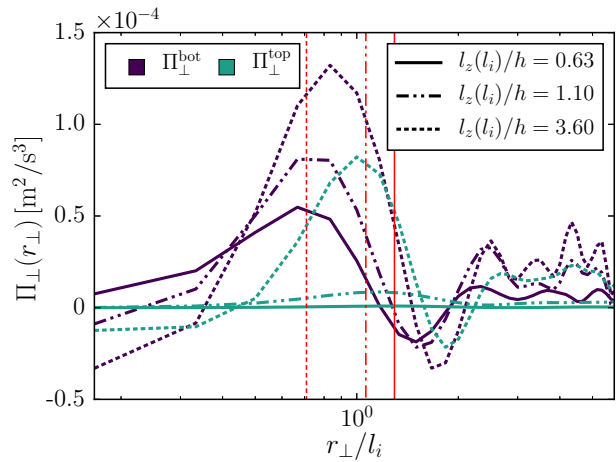

FIG. 5: Horizontal transfers of horizontal turbulent kinetic energy along the top and bottom walls, showing an inverse cascade ($\Pi_\perp > 0$) at large scales. Counter-intuitively, two-dimensionalization *promotes* a direct cascade ($\Pi_\perp < 0$) at smaller scales. Red lines locate $\hat{l}_\perp^c / l_i$.

Invoking axisymmetry, $\Pi(\mathbf{r})$ becomes a function of $r_\perp$ and $r_\parallel$ only, and splits into the four contributions[8,9]

$$\Pi_\alpha^\beta = \nabla_\alpha \cdot \langle |\delta \mathbf{u}_\beta|^2 \, \delta \mathbf{u}_\alpha \rangle, \quad (5)$$

where $\alpha$ and $\beta$ independently represent $\perp$ or $\parallel$, $\nabla_\perp \cdot = (1/r_\perp) \partial_{r_\perp}(r_\perp \cdot)$, and $\nabla_\parallel \cdot = (\partial_{r_\parallel} \cdot) \cdot \mathbf{e}_z$. None of $\Pi_\perp^\parallel$, $\Pi_\parallel^\perp$, or $\Pi_\parallel^\parallel$ can be precisely obtained from our measurements. Estimates for all contributions may nevertheless be computed as $\widehat{\Pi}_\alpha^\beta = \langle \mathbf{u}_\beta^2 \rangle \sqrt{\langle \mathbf{u}_\alpha^2 \rangle} / l_\alpha$. Fig. 4 shows all $\widehat{\Pi}_\alpha^\beta$ against $l_z(l_i)/h$. The normalization involves $l_z(l_i)$ and $E_0 = [\langle \mathbf{u}_\perp^2 \rangle + \langle \mathbf{u}_\parallel^2 \rangle]/2$ to account for the different energy levels. The only contribution to the energy transfers that strengthens, as the flow becomes quasi-2D is $\widehat{\Pi}_\perp^\perp$. This reflects that in quasi-2D channel flows, (i) the vertical velocity component becomes very small compared to the horizontal one[11,30], and (ii) velocity gradients along the magnetic field vanish. Consequently, any contribution to $\Pi$ involving $\delta \mathbf{u}_\parallel$ and/or $\partial_{r_\parallel}$ must dwindle with $l_z(l_i)/h$. In the quasi-2D limit (i.e. $l_z(l_i)/h \to \infty$), $\Pi_\perp^\perp$ coincides with $\Pi$. In any case, since $\Pi_\perp^\perp$ remains greater than the sum of all other contributions, whether the flow is 3D or not, $\Pi_\perp^\perp$ is representative of the total energy transfer $\Pi$.

In 3D MHD turbulence, Joule dissipation induces energy losses at all scales. The inertial range is accordingly reduced, and small-scale viscous dissipation is negligible when $N \gg 1$. Conversely, quasi-2D scales only experience dissipation through friction in the Hartmann layers if their turnover time exceeds the Hartmann friction time $\tau_H = h^2/\nu Ha$. In other words, energy is not conservatively transferred across the inertial range of MHD turbulence whether up- or downscale. Hence, the energy cascade does not necessarily incur a plateau region of constant energy flux. The sign of $\Pi_\perp$ is however enough to determine the direction of the transfer, as in Refs[3,8,9,31]. Note that none of our experiments displayed condensation into large turbulent structures, unlike other comparable studies[7,32,33]. This is due to a natural energy sink at large quasi-2D scales in the form of Hartmann friction, which always acted at an intermediate scale between the size of the forced region and that of the domain. This specific feature of the Flowcube ultimately enabled us to sustain statistically steady turbulence over long times.

Fig. 5 displays the horizontal transfer of horizontal kinetic energy between horizontal scales $\Pi_\perp^\perp(r_\perp)$, computed along the top and bottom Hartmann walls (referred to as $\Pi_\perp^{\text{top}}$ and $\Pi_\perp^{\text{bot}}$ respectively). The bulk of the transfers occurs in the range $r_\perp / l_i \geqslant 0.4$, for which $\Pi_\perp^{\text{bot}}(r_\perp)$ and $\Pi_\perp^{\text{top}}(r_\perp)$ are overall positive, thus indicating an upscale energy flux. The upper end of this region is dominated by oscillations, whose wavelength is close to $l_i$. These oscillations therefore likely result from the spatial inhomogeneities introduced by the forcing pattern, and/or by the non-random formation and breakup of forced vortex pairs[23,34,35]. The floor of these oscillations lies however well above noise level, which confirms, together with the continuous nature of the signature function, that random energy transfers add-up to an upscale flux at larger scales. Indeed, non-random energy transfers would translate into energy being exclusively localized at selected wavelengths only. This picture is consistent with DNS in periodic domains[36] showing that the energy cascade in MHD turbulence is local and inverse at large scales. Surprisingly, the inverse cascade extends well below $\hat{l}_\perp^c$, implying that 3D scales can also sustain an inverse cascade. What is more, two-dimensionality simultaneously promotes a direct cascade at the lower end of the spectrum (indicated by a negative value of $\Pi_\perp^\perp$), at scales lying below the saturation scale observed in Fig. 2. This behavior confirms the presence of irreducible three-dimensionality at the smallest scales. While it is not surprising that quasi-2D structures always undergo an inverse cascade, it is remarkable that some 3D scales do, and that the direct cascade affects a wider range of small scales, while two-dimensionality is promoted. These observations contrast with DNS of partly 2D and partly 3D turbulence[6,10,37], which feature sharp cascade inversions at the forcing scale only. This crucial difference may be explained by the presence of strong Joule dissipation in our setup[31,36], and/or the "broadband" nature of our forcing.

To conclude, this study shows that energy transfers are not simply governed by the topolological dimensionality of turbulence, but may also depend on the mechanisms promoting two-dimensionality and/or dissipation. In particular, MHD turbulence provides a remarkable example where an inverse energy cascade extends to topologically 3D scales. The link between turbulence kinematics and dynamics is therefore unlikely to be universal and calls for a new understanding.

### ACKNOWLEDGMENTS

AP aknowledges support from the Royal Society (Wolfson Research Merit Award Grant WM140032, International Exchanges Grant 140127), and invited professorhips from Grenoble-INP and Université Grenoble-Alpes. The SIMaP laboratory is part of the LabEx Tec 21 (Investissements d'Avenir - Grant ANR-11-LABX-0030).